\documentclass[10pt,a4paper,twoside]{article}
\usepackage{epsfig}
\usepackage{baltlat5}
\usepackage{wrapfig}
\begin{document}
\ \
\vspace{-0.5mm}

\setcounter{page}{1}
\vspace{-2mm}

\titleb{ORBITS OF FIVE VISUAL BINARY STARS}

\begin{authorl}
\authorb{B. Novakovi{\' c}}{}
\end{authorl}

\begin{addressl}
\addressb{1}{Astronomical Observatory, Volgina 7, 11160 Belgrade 74, Serbia}
\end{addressl}

\begin{summary}
We presented here the orbital parameters for five visual binary
stars calculated by using the new method which we named Sector
Grid Search. Orbital parameters were obtained for the following
stars: WDS 00152+2722 = ADS 195, WDS 02202+2949 = ADS 1780, WDS
11550$-$5606 = HIP 58106, WDS 16256$-$2327 = ADS 10049 and WDS
16256$-$2327 = ADS 10045. In addition, their masses, dynamical
parallaxes and ephemerides were calculated as well.
\end{summary}



\resthead{Orbits Of Five Visual Binary Stars}{B. Novakovi{\' c}}

\sectionb{1}{INTRODUCTION}

The study of binary stars is a very old branch of astronomy which
still plays an important role in a present astronomy. It has a
number of aims. Binary stars have been studied for decades to
measure accurate the stellar masses, to test the evolutionary
models and the star formation theories.

Although the Washington Double Star Catalog (WDS) (Mason et al.
2006) contains a large number of double and multiple systems (over
84000), according to the Sixth Catalog of Orbits of Visual Binary
Stars (Hartkopf \& Mason 2006) only for about 1900 systems the
orbital parameters have been calculated. It is because available
data for many systems are not sufficient for an accurate
determination of the orbital parameters. It is caused by two main
reasons. The number of observers is too small compared with the
number of systems and/or the orbital period in some systems is too
long compared with a time span since the beginning of systematic
measurements. As a result researchers have been pushing to
estimate system parameters from ever shorter arcs. The literature
on the determination of system parameters of visual binary is a
very numerous. Despite such a large number of methods, no one
method proposed so far is adequate in the case of a short arc. We
presented here a fully automated method for determination of
systems parameters from a short arc, which we named Sector Grig
Search (SGS). Furthermore, we also presented the systems
parameters of five visual binary stars calculated by using SGS
method. The orbits of the stars WDS 00152+2722, WDS 02202+2949 and
WDS 16256$-$2327 (pair AB) were previously announced by IAU
Commission 26 (2007, Inf. Circ. 162).

The dynamical parallaxes and the individual masses were calculated
for stars which belong to the Main Sequence according to the
method proposed by Angelov (1993). Trigonometric parallaxes
published in the Hipparcos and Tycho Catalogues (ESA 1997) were
used in order to calculate the total masses of the systems. The
values of the stellar masses as a function of the spectral type,
used as reference, have been taken from Schmidt-Kaler (1982).
Visual magnitudes and spectral types, presented in Table 3. were
taken from WDS catalogue.

\vskip1mm

\sectionb{2}{METHOD}

Among the methods presented so far one can find several efficient
methods. We will mention here only a few of them which are widely
used.

The best known method for determination of system parameters is
that of Thiele-Innes-van den Bos (Thiele 1883; van den Bos 1926,
1932; Dommanget 1981). Yet it cannot be expected to handle all
possible cases. It requires the knowledge of three normal places
and an area constant. Docobo (1985) proposed an analytical method
which does not require the knowledge of the area constant. The
method is based on a mapping from the interval (0,2$\pi$) into the
family of Keplerian orbits whose apparent orbits pass through
three base points. These points are taken either coinciding with
the most reliable measurements or belong to the areas with a
maximum of observational evidence in their favor, but do not
necessary coincide with actual measurements. The orbits which fits
the best to all known measurements is chosen from all generated
possible ones. In addition, it has an option allowing the use the
radial velocities, being thus useful for the inclinations close to
$90^\circ$ and in the case of a short arc as well.

The numerical methods, based on minimization procedure, which take
advantage from computer's power were proposed by Eichhorn and Xu
(1990) and Pourbaix (1994). The method proposed by Hartkopf et
al.(1989) also takes advantage from computer power but with a
different approach. The orbital parameters are calculated
according to the "three-dimensional adaptive grid search" and it
requires preliminary knowledge of the period $P$, epoch of
periastron passage $T$ and eccentricity $e$.

The most recently Olevi{\'c} \& Cvetkovi{\'c} (2004) proposed a
method suitable for application in the case of short arc which is
based on using supplementary, fictive, measurements. In our
previous papers (e.g. Novakovi{\'c} 2007) we used this method in
order to calculate the orbital parameters and we obtained
acceptable results when deal with a short arc. Although we
obtained good results with this method, they depend on the choice
of the fictive measurements.

We presented here a simple and fully automated method which can
give a reliable orbit solution even in the case of a short arc and
does not need any subjective initial parameters such as fictive
measurements. It consists of several main steps.

The first step in our procedure is to assign the appropriate
weights to all measurements according to the weighting rules
described in Hartkopf et al. (1989,2001) and rejecting all
measurements with errors which do not satisfied $3\sigma$
condition. The next step is to divide observed arc in five equal
intervals and determining minimum and maximum value of angle
separation for every interval $(\rho_{min}, \rho_{max})_{i=1,5}$
as well as an average value of position angle for every interval
$\theta_{i=1,5}$. \footnote{In order to include the possible
measurements errors it is necessary to correct the values of
$\rho_{min}$ and $\rho_{max}$ (correction makes interval a little
bit wider). This correction should takes into account a maximum
allowed error of $3\sigma$.} Given the set of the five position
angles $\theta_{i=1,5}$ and the five intervals of the angle
separations $(\rho_{min}, \rho_{max})_{i=1,5}$ we can proceed
further. As we know a general equation of the conic section is of
the form

\vskip5mm

\begin{equation}
\label{eq1}
 a_{20}x^{2}+a_{02}y^{2}+2a_{11}xy+2a_{10}x+2a_{01}y-1 = 0
\end{equation}

\vskip3mm

so the curve has been determined if we have determined the five
coefficients $a_{20}$,$a_{02}$,$a_{11}$,$a_{10}$,$a_{01}$. If the
positions of the five points were given the coefficients can be
determined by solving system of the five linear equations. We
apply an iterative procedure beginning with a $\rho_{min}$ and
increasing the angle separation $\rho$ for step size as many times
as we need to reach $\rho_{max}$. For every combination of the
five position angles and the five angle separations the five
coefficients are determined by using LU decomposition described in
Press et al. (1992). From these five coefficients we can calculate
the five orbital parameters ($a,e,i, \Omega, \omega$) and two
dynamical parameters ($P,T$) can be calculated from the Keplerian
equation

\vskip3mm

\begin{equation}
\label{eq2}
\frac{2\pi}{P}(t_{j}-T)=M(\theta_{j},a,e,i,\Omega,\omega).
\end{equation}
The last step is minimization of function D defined as
\begin{equation}
\label{eq3} D=[\sum_{j} w_{j}((x_{o}-x_{c})^2_{j} +
(y_{o}-y_{c})^2_{j})]/\sum_{j} w_{j}.
\end{equation}

In this formula, $w_{j}$  denotes the weight of the $j^{th}$
measurement, $x_{o}$,$y_{o}$ denotes the observed positions and
$x_{c}$,$y_{c}$ denotes the calculated positions of the companion.
The minimization is done by applying Powell's method (Press et al.
1992), but with just three degrees of freedom ($P,T,a$) in order
to decrease probability to reach a local minimum which number
growing as $\exp(N)$ ($N$-the number of degrees of freedom). For
each step in the grid search a set of orbital parameters is
calculated and a solution with minimum residuals is accepted as a
final solution.

\subsectionb{2.1}{Test}

In order to test efficiency of SGS method we determined the
orbital parameters of two visual binary stars which have had
orbital parameters already calculated and very reliable. For this
test we selected the star WDS 02022+3643 (HD 12376) from the
Catalog of Orbits and Ephemerides of Visual Double Stars (Docobo
et al. 2001) and the star 16413+3136 (HD 150680) from the Sixth
Catalog of Orbits of Visual Binary Stars. In the case of the star
WDS 02022+3643 test was made using speckle measurements only, but
in the case of the star WDS 16413+3136 test results were obtained
using mixed data (visual,speckle,photographic,CCD).

The star WDS 02022+3643 (pair AB) has a very good orbit calculated
by Hartkopf et al. (2000). As we wanted to show that method
presented here is efficient in the case of a short arc we did not
use all measurements available for this star but only a set of 7
measurements from 1994.7086 to 1999.8856 which covered an arc of
$\approx$ $59^\circ$. Similarly, we calculated orbital parameters
for binary star WDS 16413+3136 using only a part of available
measurements. For the purpose of this study we chose 38
measurements from 1989.3931 to 1999.61 which covered an arc of
$\approx$ $67^\circ$.

Test results are presented in a Table 1. and they show that by
using SGS method good results can be obtained even in the case of
a short arc.

\begin{table}
\caption{Test results} \label{T1} \vskip 2mm
\begin{tabular}{|cc|cc|cc|}\hline
Element & &SGS method & Hartkopf et al. 2000  & SGS method&Soderhjelm 1999 \\
\hline
$P[yr]$& &12.90& 12.94 &36.76 & 34.45\\
$T$& &1988.82&1989.06  &1966.1 & 1967.7 \\
$a[^{\prime\prime}]$ & &0.155&0.150  &1.40 &1.33 \\
$e$ & &0.508& 0.404 & 0.40& 0.46\\
$i[^o]$& &68.9&67.0  &127.4 & 131.0\\
$\Omega[^o]$& &195.2&191.4  & 53.7& 50.0\\
$\omega[^o]$& &289.0& 295.1 &107.4 & 111. 0\\
\hline
\end{tabular}
\end{table}

\sectionb{3}{RESULTS AND COMMENTS}

In this section we presented results and comments on individual
objects.

Table 2. presents the corresponding numerical values for the
orbital parameters (epoch J2000) and their estimated formal errors
as well as the identifications of the stars in several widely-used
catalogs.

Table 3. gives the astrophysical quantities for both components:
visual magnitudes, spectral types, absolute visual magnitudes,
masses and, in the last two columns, the calculated dynamical
parallax and the Hipparcos trigonometric parallax.

Table 4. gives predicted ephemerides for these systems for the
period 2008-2012.

The orbits are illustrated in Figures 1-5. The solid curves
represent the newly determined orbital parameters, while the
dashed curves represent previously published orbital parameters.
The solid lines indicate the line of nodes. All measurements are
connected with their predicted positions on the new orbit by "O-C"
lines. The interferometric measurements are represented of each
plot by fulled circles "$\bullet$" and all others measurements
(visual, photographic) are represented by plus "+". The direction
of motion is indicated on the north-east orientation in the lower
right of each plot and "+" indicate position of primary star.

\textbf{WDS 00152+2722 = J 868}. It was discovered in 1907.78 by
Barton (1926) and 15 measurements have been performed till a year
2002. Recently, an additional measurement was carried out by our
double star group (Cvetkovi{\'c} et al. 2007). Our orbital
solution is the first one for this binary star and it was
calculated by using set of all 16 measurements. As data about this
star are very poor we do not know very much about it. Under the
assumption that its components belong to the Main Sequence we
estimated the total mass of system ${\cal M}_{tot}$ $\approx$ 1.0
${\cal M}_{\odot}$.

\textbf{WDS 02202+2949 = A 961}. This binary star was discovered
in 1905.87 by Aitken (1906) at Lick Observatory. Previous orbit of
this binary have been calculated by Heintz (1969). In order to
demonstrate improvement of previous orbit obtained by Heintz we
calculated root mean square (rms) of residual values\footnote{When
deriving these values corresponding weights have been taken into
account}. The obtained rms values are $\Delta\theta$ = 2.9$^{o}$,
$\Delta\rho$ = 0\farcs039 (this work), and $\Delta\theta$ =
2.9$^{o}$, $\Delta\rho$ = 0\farcs042 (Heintz's orbit) what
indicates that our orbit solution does improve separation but
gives statistically same result for position angle. Our orbital
parameters and the Hipparcos parallax yield a total mass of system
${\cal M}_{tot}$ $\approx$ 3.0 ${\cal M}_{\odot}$. This value is
slightly larger then expected value for this spectral type (on
condition that both components belong to the Main Sequence).

\textbf{WDS 11550$-$5606 = HLD 114}. This is a wide binary star
discovered in 1882 by Holden (1884). Till the present time 47
measurements have been performed and they cover an arc of
$\approx$60$^{o}$. Our orbital parameters are the first ones
calculated for this binary and they should be classified as
preliminary. The new measurements are necessary to obtain more
reliable orbit solution and to understand significant difference
between dynamical parallax (this work) and Hipparcos trigonometric
parallax. In general, this disagreement could be due to the
uncorrect orbital solution and/or because at least one of the
components does not belong to the Main Sequence. The spectral type
of primary star (G3IV/V) suggested that it could lies outside of
the Main Sequence.

\textbf{WDS 16256$-$2327 = H 2  19 AB}. It was discovered as early
as in 1780.34 by W. Herschel. Though the time interval after the
discovery has already exceeded 200 years, available measurements
cover a short arc of $\approx$ $28^\circ$. This pair is included
in the Catalog of Rectilinear Elements (Hartkopf et al. 2006)
indicated that it is not a physical. Despite this indication, we
calculated system parameters because we think that is not clear
yet whether this pair is physical or optical. The rms values are
$\Delta\theta$ = 1.9$^{o}$, $\Delta\rho$ = 0\farcs147 (this work),
and $\Delta\theta$ = 1.8$^{o}$, $\Delta\rho$ = 0\farcs148 (linear
fit) what is statistically indistinguishable results. Our orbital
parameters and Hipparcos parallax yield a total mass of the system
${\cal M}_{tot}$ $\approx$ 23.6 ${\cal M}_{\odot}$. As one can
expect near 10.0${\cal M}_{\odot}$ (Schmidt-Kaler 1982) for stars
of B2V spectral type and slightly over 10.0${\cal M}_{\odot}$ for
B2IV spectral type, the obtained value of total mass of the system
is reasonable.

\textbf{WDS 16256$-$2327 = BU 1115 DE}. This binary star belong to
the same multiple system as the previous one, but it was
discovered by Burnham (1894) in 1889.39 at Lick Observatory. The
orbital parameters presented here are the first for this binary.
The dynamical parallax is $\pi = 0\farcs00769$, in very good
agreement with the Hipparcos parallax $\pi_{HIP} = 0\farcs00733$,
but the total mass of the system (${\cal M}_{tot}$ $\approx$ 5.8
${\cal M}_{\odot}$ with Hipparcos parallax), calculated with our
orbital parameters and either Hipparcos or dynamical parallax, is
significantly smaller then expected for B5V spectral type. As we
know, combination of the uncertainty of the semi-major axis, of
the orbital period and that of the parallax is easily propagated
in the standard deviation of the total mass as,

\begin{equation}
  \label{eq4}
\sigma_{M}={\cal
M}_{tot}\sqrt{9(\frac{\sigma_{a}}{a})^{2}+9(\frac{\sigma_{\pi}}{\pi})^{2}+4(\frac{\sigma_{P}}{P})^{2}}
\end{equation}
According to equation \,(\ref{eq4}) the error of total mass is
$\sigma_{M} = 4.0$ ${\cal M}_{\odot}$. Taking into account this
error, we obtained 9.8 ${\cal M}_{\odot}$ as the upper limit for
the total mass of this system.

\begin{figure}
\begin{minipage}[b]{0.5\linewidth} 
\centering
\includegraphics[width=80mm]{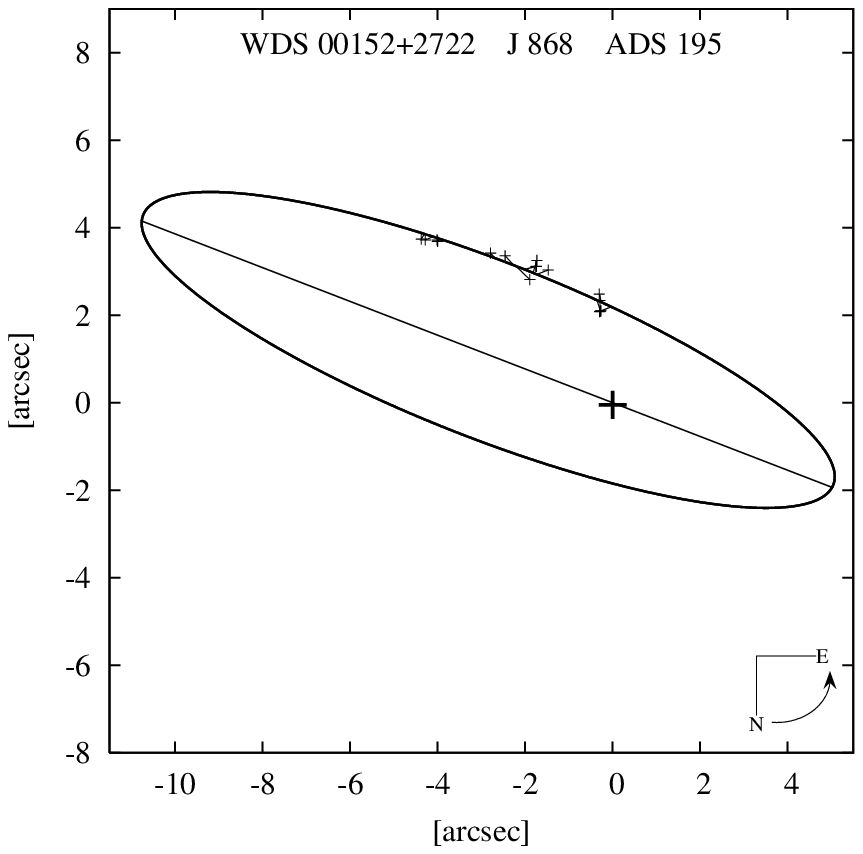}
\caption{J 868.}
\end{minipage}
\hspace{0.5cm} 
\begin{minipage}[b]{0.5\linewidth}
\centering
\includegraphics[width=80mm]{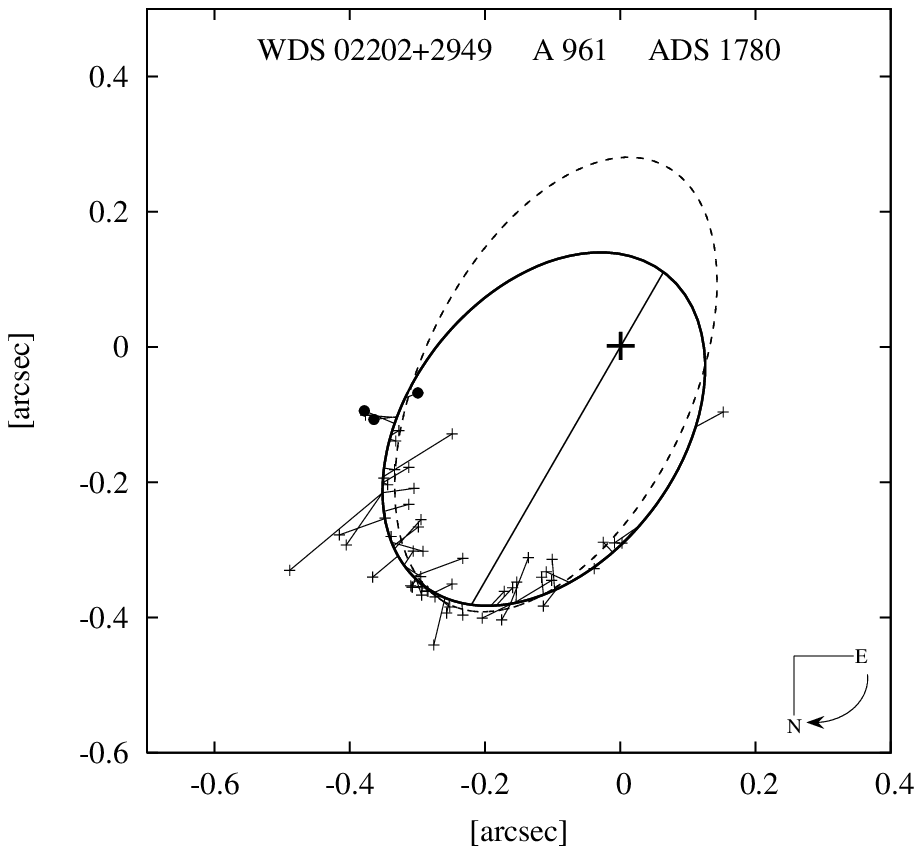}
\caption{A 961.}
\end{minipage}
\end{figure}

\begin{figure}
\begin{minipage}[b]{0.5\linewidth}
\centering
\includegraphics[width=80mm]{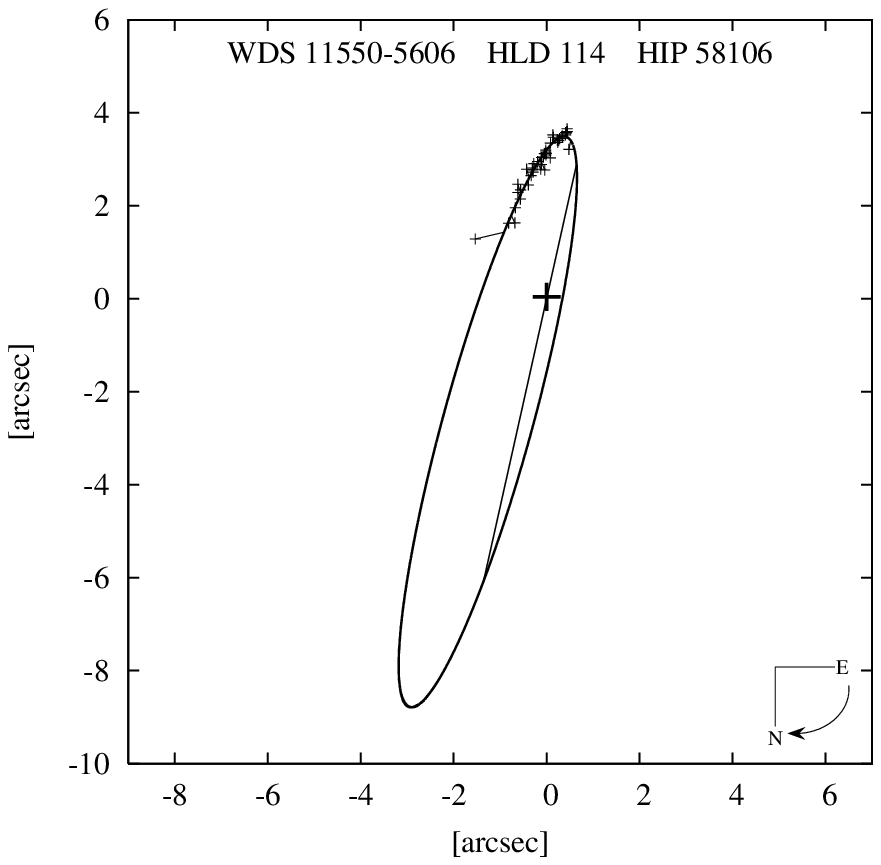}
\caption{HLD 114.}
\end{minipage}
\hspace{0.5cm}
\begin{minipage}[b]{0.5\linewidth}
\centering
\includegraphics[width=80mm]{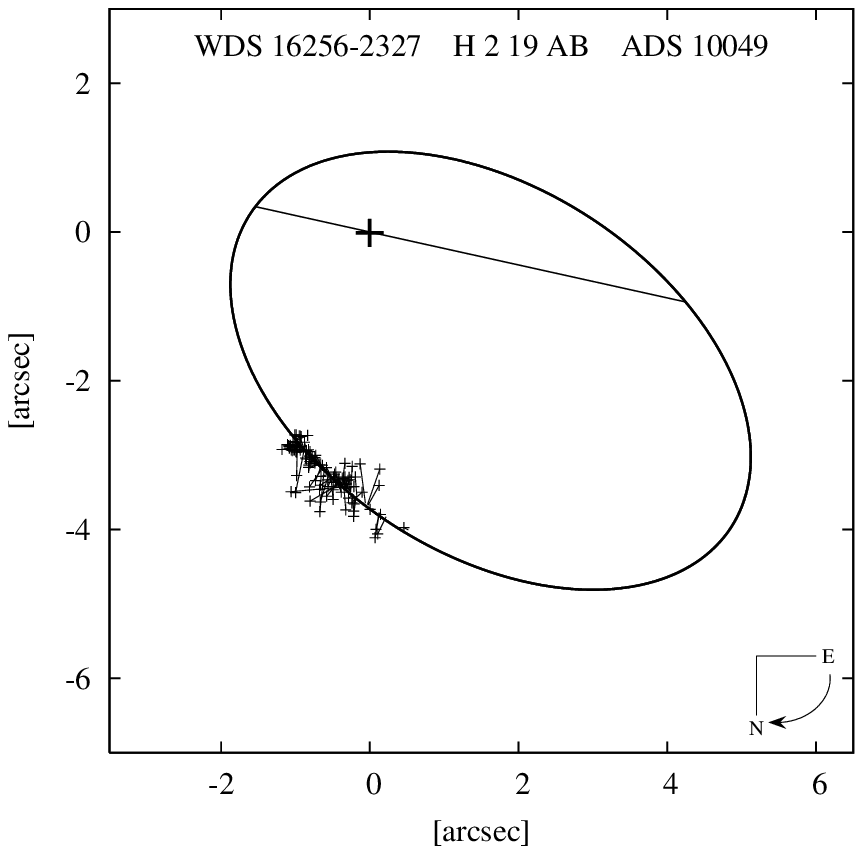}
\caption{H 2  19 AB.}
\end{minipage}
\end{figure}

\vskip5mm

\begin{center}
\vbox{\small \tabcolsep=5pt
\begin{tabular}{cccccccccc}
\multicolumn{10}{c}{\parbox{90mm}{
{\normbf \ \ Table 2.}{\norm\ Orbital parameters (J2000).}}}\\
\tablerule
Name   &ADS &HD   &              &              &            &            &             &             &                \\
WDS &BDS &HIP &$P[yr]$ & $T$ & $a[^{\prime\prime}]$ & $e$ & $i[^o]$ & $\Omega[^o]$ & $\omega[^o]$ \\
\tablerule
            &&&               &            &             &             &            &           &          \\
J 868       &195  &-&    1089. &    1725. &      8.48  &      0.367  &     75.2   &    68.9   &    352.8 \\
00152+2722  &-&-&   $\pm$153. &  $\pm$147.&  $\pm$1.06 &  $\pm$0.370 &  $\pm$2.5  & $\pm$3.1 & $\pm$19.4  \\
            &&&               &            &             &             &            &           &          \\
A 961       &1780 &14394&      143.1   &    2038.1 &      0.294  &      0.571  &    145.4   &   150.0   &    15.6  \\
02202+2949  &12885&10892&  $\pm$5.6 & $\pm$5.7& $\pm$0.057 & $\pm$0.095 & $\pm$3.9 & $\pm$5.8 & $\pm$8.5 \\
            &&&               &            &             &             &            &           &          \\
HLD 114     &-&103493&  930.   & 2039. & 8.02  & 0.707  & 97.7  &  167.4  & 59.7  \\
11550$-$5606  &-&58106&   $\pm$98.  & $\pm$91.& $\pm$0.90 & $\pm$0.109 & $\pm$1.5 & $\pm$1.3 &$\pm$7.4\\
            &&&               &            &             &             &            &           &          \\
H 2  19 AB  &10049&147933&    2398.   &    2327. &      4.25  &      0.675  &     135.3  &    77.5   &    226.1 \\
16256$-$2327  &7613&80473&  $\pm$326.&  $\pm$343.&  $\pm$0.79& $\pm$0.322 &  $\pm$6.9 & $\pm$13.5 &$\pm$15.3 \\
            &&&               &            &             &             &            &           &          \\
BU 1115 DE  &10045&147888& 675.5   & 2008.6  & 1.01   & 0.707   & 134.8    & 152.7  & 260.4    \\
16256$-$2327  &7609&80461&  $\pm$32.5  &  $\pm$34.2&  $\pm$0.15 & $\pm$0.112  &  $\pm$2.7 & $\pm$4.6 &$\pm$1.1  \\
\hline
\end{tabular}
}
\end{center}

\vskip8mm

\begin{center}
\vbox{\small \tabcolsep=5pt
\begin{tabular}{lccccccccc}
\multicolumn{10}{c}{\parbox{90mm}{
{\normbf \ \ Table 3.}{\norm\ Dynamical elements.}}}\\
\tablerule
WDS  &    $m_A$--$m_B$   &  Sp.  &  $M_A$   & $M_B$ &  ${\cal M}_A $ &  ${\cal M}_B$& $\pi_{dyn}$ & $\pi_{HIP}$ $\pm$ $\sigma_{\pi_{HIP}}$ \\
     &                   &       &          &       &                &  &  &  &  \\
\tablerule
            &              &            &          &          &          &         &         &              \\
00152+2722  &12.42 - 12.62 &      -     &  11.91   &  12.21   &   0.54   &  0.50   &  79.06  &     -       \\
            &              &            &          &          &          &         &         &             \\
02202+2949  & 9.31 - 8.94  &    F5      &   3.46   &   3.09   &   1.33   &  1.41   &   7.69  &7.48 $\pm$ 2.59\\
            &              &            &          &          &          &         &         &             \\
11550$-$5606  & 7.36 - 7.81&   G3IV/V   &   6.18   &  6.63    &   0.88   &  0.82   &  70.50  &32.33 $\pm$ 1.44\\
            &              &            &          &          &          &         &         &             \\
16256$-$2327(AB)&5.07  5.74&  B2IV B2V  &    -     &    -     &    -     &   -     &   -    &8.27 $\pm$ 1.18\\
            &              &            &          &          &          &         &         &              \\
16256$-$2327(DE)& 7.04  8.65&    B5V    &   1.47   &   3.08   &   3.06   &  1.97   &   7.69  &7.33 $\pm$ 1.37\\
\hline
\end{tabular}
}
\end{center}

\vskip8mm

\begin{center}
\vbox{\small \tabcolsep=5pt
\begin{tabular}{ccccccc}
\multicolumn{7}{c}{\parbox{90mm}{
{\normbf \ \ Table 4.}{\norm\ Ephemerides.}}}\\
\tablerule
   &  2008 & 2009 & 2010 & 2011 & 2012 \\
   &       &      &      &      &      \\
 WDS \quad Desig. \ \ \ \ &
 $\theta $ \ \quad $\rho $ &
 $\theta $ \ \quad $\rho $ \ \ &
 $\theta $ \ \quad $\rho $ &
 $\theta $ \ \quad $\rho $ &
 $\theta $ \ \quad $\rho $ \\
 $\alpha,\delta \ (2000) $ \ \ \ \ &
 $[^o] $ \ $[^{\prime\prime}]$ &
 $[^o] $ \ $[^{\prime\prime}]$ \ \ &
 $[^o] $ \ $[^{\prime\prime}]$ &
 $[^o] $ \ $[^{\prime\prime}]$ &
 $[^o] $ \ $[^{\prime\prime}]$ \\
\tablerule
    &     &      & &      &      &      \\
 00152+2722......& 228.3 \ \ 5.817 & 228.5 \ \ 5.855 & 228.6 \ \ 5.892 & 228.8 \ \ 5.929 & 229.0 \ \ 5.967\\
 02202+2949......& 277.4 \ \ 0.301 & 275.7 \ \ 0.294 & 274.0 \ \ 0.287 & 272.1 \ \ 0.280 & 274.2 \ \ 0.273\\
 11550$-$5606......& 169.7 \ \ 3.291 & 169.4 \ \ 3.264 & 169.2 \ \ 3.235 & 169.0 \ \ 3.205 & 168.7 \ \ 3.172 \\
 16256$-$2327(AB)..& 338.6 \ \ 2.913 & 338.5 \ \ 2.908 & 338.3 \ \ 2.903 & 338.1 \ \ 2.898 & 337.9 \ \ 2.893\\
 16256$-$2327(DE)..& 259.9 \ \ 0.214 & 253.8 \ \ 0.211 & 247.7 \ \ 0.210 & 241.5 \ \ 0.210 & 235.4 \ \ 0.213 \\
\hline
\end{tabular}
}
\end{center}

\vskip5mm

\begin{figure}
\begin{minipage}[b]{0.5\linewidth}
\centering
\includegraphics[width=80mm]{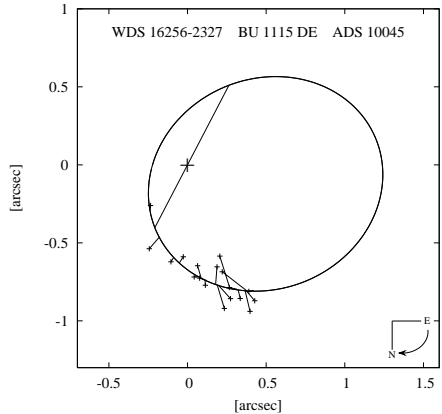}
\caption{BU 1115 DE.}
\end{minipage}
\end{figure}

\vskip 3mm

\sectionb{4}{CONCLUSIONS}

We presented here a new method for determination of the orbital
parameters of visual binary stars. This method yields to the
reliable orbit solution even in the case of a short arc. On the
other hand SGS method is fully automated and very easy to use what
is specially important when deal with a large amount of data. Thus
the method will be useful in future astrometric missions such as
SIM\footnote{Space Interferometry Mission (SIM),
http://sim.jpl.nasa.gov/} and GAIA\footnote{Global Astrometric
Interferometer for Astrophysics (GAIA),
http://astro.estec.esa.nl/GAIA/}. In order to improve the orbital
parameters of visual binaries presented here, new measurements are
necessary. The observers should pay the more attention to the star
BU 1115 DE which will pass through the periastron at the middle of
2008.

\vskip5mm

ACKNOWLEDGMENTS. I would like to express my thanks to Dr.
J.A.Docobo (referee) and Dr. Z. Cvetkovi{\'c} for their useful
comments and suggestions. This research have made use of the
Washington Double Star Catalog maintained at the U.S. Naval
Observatory and the Simbad database operated at CDS, Strasbourg,
France. This research has been supported by the Ministry of
Science of the Republic of Serbia (Project No 146004 "Dynamics of
Celestial Bodies, Systems and Populations").

\vskip5mm

\References

\vskip2mm

\refb Aitken~R.~G. 1906, Lick Obs. Bull. 4, 4

\refb Angelov~T. 1993, Bull. Astron. Belgrade, 148, 1

\refb Barton~S.~G. 1926, AJ, 36, 155

\refb Burnham~S.~W. 1894, Publ. Lick Obs. 2, 1

\refb Cvetkovi{\'c}~Z., Pavlovi{\'c}~R., Strigachev~A.,
Novakovi{\'c}~B., Popovi{\'c}~G.~M. 2007, Serb. Astron. J., 174,
83

\refb Docobo~J.~A. 1985, Celest. Mech., 36, 143

\refb Docobo~J.~A., Ling~J.~F., Prieto~C., Costa~J.~M.,
Costado~M.~T. and Magdalena~P. 2001, Acta Astronomica 51, 353

\refb Dommanget~J. 1981, A\&A, 94, 45

\refb Eichhorn~K.~H., Xu~Yu-Lin 1990, AJ, 358, 575

\refb ESA 1997, The Hipparcos and Tycho Catalogues, ESA SP-1200

\refb Hartkopf~W.~I., McAlister~H.~A., Franz~O.~G. 1989, AJ, 98,
1014

\refb Hartkopf~W.~I., Mason~B.~D., McAlister~H.~A. 2000, AJ, 119,
3084

\refb Hartkopf~W.~I., Mason~B.~D., Worley~C.~E. 2001, AJ 122, 3472

\refb Hartkopf~W.~I., Mason~B.~D. 2006, Sixth Catalog of Orbits of
Visual Binary Stars, US Naval Observatory, Washington. Electronic
version http://ad.usno.navy.mil/wds/orb6.html

\refb Hartkopf~W.~I., Mason~B.~D., Wycoff~G.~L., Kang~D. 2006,
Catalog of Rectilinear Elements, U.S. Naval Observatory,
Washington. Electronic version
http://ad.usno.navy.mil/wds/lin1.html

\refb Heintz~W.~D. 1969, A\&A, 2, 169

\refb Holden~E.~S. 1884, Publ. Washburn Obs. 2, 97

\refb Mason~B.~D., Wycoff~G.~L., Hartkopf~W.~I. 2006, The
Washington Double Star Catalogue, US Naval Observatory,
Washington. Electronic version
http://ad.usno.navy.mil/wds/wds.html

\refb Novakovi{\' c}~B. 2007, Chin. J. Astron. Astrophys. 7, 415

\refb Novakovi{\' c}~B. 2007, IAU Commission 26 Inf. Circ., 162

\refb Olevi{\'c}~D., Cvetkovi{\'c}~Z. 2004, A\&A, 415, 259

\refb Pourbaix~D. 1994, A\&A, 290, 682

\refb Press~W.~H., Teukolsky~S.~A., Vettering~W.~T.,
Flannery~B.~P. 1992, in {\it Numerical recipes in Fortran. The art
of scientific computing.}, 2nd ed., Cambridge University Press,
New York

\refb Schmidt-Kaler~Th. 1982, in: {\it Numerical Data and
Functional Relationships in Science and Technology – New Series},
eds. L.~H.~Aller, I.~Appenzeller, B.~Baschek et al.,
Landolt-B¨ornstein 2-B, p. 1

\refb Soderhjelm~S. 1999, A\&A, 341, 121

\refb Thiele~T.~N. 1883, Astron. Nachr., 104, 245

\refb van den Bos~W.~H. 1926, Union Circ., 68, 354

\refb van den Bos~W.~H. 1932, Union Circ., 86, 261

\end{document}